\documentclass[review]{elsarticle}

\usepackage{hyperref}

\usepackage{amssymb}
\usepackage{amsmath}
\usepackage{multicol}
\usepackage{booktabs}

\usepackage[figuresright]{rotating}
\usepackage{bm}
\usepackage{caption}

\begin{document}

\begin{frontmatter}

\title{Privacy-Preserving Distributed Machine Learning Made Faster}

\author{Zoe L. Jiang, Jiajing Gu, Hongxiao Wang, Yulin Wu, Junbin Fang, Siu-Ming Yiu, Wenjian Luo, Xuan Wang}

%
%

\begin{abstract}
With the development of machine learning, it is difficult for a single server to process all the data. So machine learning tasks need to be spread across multiple servers, turning the centralized machine learning into a distributed one. However, privacy remains an unsolved problem in distributed machine learning. Multi-key homomorphic encryption is one of the good candidates to solve the problem. However, the most recent result about Multi-key homomorphic encryption scheme (MKTFHE) only supports NAND gate. Although it is Turing complete, it requires efficient encapsulation of NAND gate to further support mathematical calculation. This paper designs and implements a series of operations on positive and negative integers accurately. First, we design basic bootstrapped gates with the same efficiency as that of the NAND gate. Second, we construct practical $k$-bit complement mathematical operators based on our basic binary bootstrapped gates. The constructed created can perform addition, subtraction, multiplication and division on both positive and negative integers. Finally we demonstrated the generality of the designed operators by achieving a distributed privacy-preserving machine learning algorithm, i.e. linear regression with two different solutions. Experiments show that the operators we designed are practical and efficient.
\end{abstract}


\end{frontmatter}

\section{Introduction}
In the big data era, it is necessary to spread the machine learning tasks across multiple servers and transform centralized systems into distributed ones~\cite{verbraeken2020survey}. These distributed systems present new challenges and especially privacy remains an unsolved challenge in it~\cite{PMP4MLDS18}.

Privacy computing is a kind of techniques which perform data computation without specified information leakage. Homomorphic encryption is a special form of encryption that permits users to perform computations on encrypted data without first decrypting it, which has great practical implications in the outsourcing of private computations. It is called fully homomorphic encryption (FHE) if both addition and multiplication are supported. The security of FHE is usually guaranteed by the hard problems on lattices, which is also considered as one of the post-quantum cryptographic schemes~~\cite{post-quantum}. However, computing on encrypted data using such FHE schemes will increase the noise in the ciphertext. If the noise in the ciphertext grows beyond some threshold, it will result in wrong decryption~\cite{CrawfordGHPS18}. This problem can be solved by a bootstrapping technique, first proposed by Gentry~\cite{Gentry09}. Bootstrapping can reduce the noise in ciphertext to refresh it so that the homomorphic evaluation can be performed constantly.

Obviously, \emph{single-key} fully homomorphic encryption only allows server to perform addition and multiplication on data encrypted by same key. \emph{Multi-key} fully homomorphic encryption (MKFHE) was proposed to circumvent this shortcoming~\cite{Lopez12}. It enables users to encrypt their own data under their own keys, while homomorphic evaluations can be performed on encrypted data directly at server side without decryption. It avoids the possibility that user and server conspire to steal the data of other users. Therefore, MKFHE indeed realizes secure multi-party computation with untrusted party. While Chen et al.~\cite{MKTFHE} developed the library Multi-key fully homomorphic encryption over torus (MKTFHE) for NAND gate.

However, MKTFHE library only provides multi-key homomorphic NAND gates. Perform multi-key homomorphic computation using only homomorphic NAND gates is a complex and time-consuming work. As a result, a simple multi-key homomorphic NAND gate is not practical to be used directly. In order to better apply multi-key homomorphic encryption on privacy computing, we design and implement a series of practical multi-key homomorphic mathematical operators.

In this study, we make the following contributions :

\begin{enumerate}
	\item We designed a series of fundamental multi-key bootstrapped gates based on MKTFHE, including multi-key bootstrapped AND gate, multi-key bootstrapped OR gate, multi-key bootstrapped NOR gate, multi-key bootstrapped XOR gate, multi-key bootstrapped XNOR gate, and multi-key NOT gate (without gate bootstrapping). Experiment results show that the proposed fundemental multi-key bootstrapped gates are more efficient than directly joining multi-key bootstrapped NAND gates to build basic binary gates.
	
	\item We designed a series of multi-key homomorphic operators based on the basic multi-key bootstrapped gates we constructed. The operators we designed include $k$-bit complement array integer adder, $k$-bit complement array integer subtractor, $k$-bit complement array integer multiplier and $k$-bit complement array integer divider, which can perform addition, subtraction, multiplication and division on arbitrary bits integers in both positive and negative. We construct the complement array adder linearly and used a structure similar to the 4-bit array integer multiplier in~\cite{Hwang79} to construct our $k$-bit complement multiplier. In the similar way we construct the $k$-bit complement array integer divider. The subtractor is constructed with a adder and a simple multiplier. The time of the adder and the subtractor grows linearly with the bits of input numbers, while the time of the multiplier and divider grows quadratically. The time of the divider grows almost linearly with the number of layers in divider.
	
	\item We train linear regression model by utilizing our proposed multi-key homomorphic operators. Taking linear regression as an example, we implement a whole multi-key fully homomorphic machine learning scheme. Experimental results show that the practicability and generality of our homomorphic operator and the training time grows linearly with the number of participants. 
\end{enumerate}


The rest of this paper is organized as follows. Section \ref{2Rel} discusses the related research work. In Section \ref{2Pre}, it clarifies the notation and review some constructions related. Section \ref{3Scheme} describes the $k$-bit complement array operators with implementation based on MKTFHE and the distributed machine learning model we evaluate, followed by the performance analysis and experimental results of our implementation in Section \ref{4Imp}. Section \ref{5Con} concludes the paper with future work.

\section{Related Work}

\label{2Rel}

Previous research efforts have made a number of contributions to the development of privacy-preserving machine learning. Some work has focused on securely outsourcing the training of ML models to the cloud, typically by using homomorphic encryption techniques~\cite{froelicher2020scalable}. In 2016, Aono et al.~\cite{aono2016scalable} proposed a secure system for protecting the training data in logistic regression via homomorphic encryption. In 2018, Crawford et al.~\cite{crawford2018doing} built a system that uses fully-homomorphic encryption to approximate the coefficients of a logistic-regression model built from genomic data, while Kim et al.~\cite{kim2018logistic, kim2018secure} presented a method to train a logistic regression model without information leakage. However, these scheme are all based on simple-key fully homomorphic encryption, which are not suitable for distributed machine learning models.

In 2012, Graepel et al.~\cite{GraepelLN12} first proposed that it is possible to perform machine learning algorithm on encrypted data by homomorphic encryption scheme. Then, several single-key homomorphic encryption schemes are used for machine learning prediction~\cite{CryptoNets16, BourseMMP18, BoemerLCW19, TianNYY21, CHET19} or machine learning training~\cite{ChenGHHJLL18, KimS0XJ18, CheonKKS18}. A few machine learning algorithms has adapted multi-key homomorphic encryption and demonstrated the premising future~\cite{DKS19}.

Multi-key fully homomorphic encryption was first proposed based on the NTRU assumption by Lopez et al.~\cite{Lopez12} at 2012, which is intended to apply to on-the-fly multiparty computation. In 2015, Clear et al.~\cite{Clear15} constructed the first LWE-based MKFHE scheme, and improved by Mukherjee and Wichs~\cite{Mukherjee16} in 2016. These schemes are single-hop MKFHE schemes, which means all the participants must be known in advance. This problem was solved by Peikert et al.~\cite{Peikert16} and Brakerski et al.~\cite{Bra16} in 2016 by constructing multi-hop MKFHE schemes. However, these schemes are impractical and without implementation. 

The first implementation of the MKFHE scheme was achieved by Chen et al.~\cite{MKTFHE} in 2019, named MKTFHE. In their scheme, a multi-key bootstrapped NAND gate is described. This scheme was improved by Lee and Park~\cite{LeeP19} in 2019, realizing the distributed decryption in MKTFHE. However, using multi-key bootstrapped NAND gates to construct homomorphic encryption operation directly has disadvantages such as low practicability, high complexity and error-prone construction process. At the same time, the complex structure lead to expensive optimization.

\section{Preliminaries}

\label{2Pre}
\subsection{Notation}

The rest of the paper uses the following notations. $\mathbb{T}$ denotes the real Torus $\mathbb{R}/\mathbb{Z}$, the set of real numbers modulo $1$. $\mathbb{T}_N [X]$ denotes $\mathbb{R}[X]/(X^N + 1)$ mod $1$. $k$ represents the number of parties. TLWE is used to denote the (scalar) binary learning with error problem over Torus, while for the ring mode, we use the notation of TRLWE. $params$ represents the parameter sets used in TFHE (fully homomorphic encryption over torus) scheme, while $mkparams$ represents the parameter sets used in MKTFHE scheme and our scheme.

\subsection{MKTFHE}

MKTFHE scheme is the multi-key version of TFHE scheme. TFHE, constructed by Chillotti et al.~\cite{ChillottiGGI16, ChillottiGGI17, ChillottiGGI20}, is a fast fully homomorphic encryption (FHE) scheme over the torus, which generalizes and improves the FHE based on GSW~\cite{GentrySW13} and its ring variants. In the TFHE scheme, bootstrapped binary gates are designed to represent the functions developers need.

The main idea of TFHE is to bootstrap after every binary gate evaluation to refresh the ciphertext in order to make it usable for the following operations, resulting in that arbitrarily deep circuits can be homomorphically evaluated. That is to say, a single parameter set allows the sever to evaluate any function. The entire homomorphic evaluation of the circuits will take time proportional to the number of the binary gates used or, if parallelism is involved, to the number of the circuit layers.

The message space of TFHE bootstrapping gates is $\mathbb{T}$. A TLWE ciphertext $(a, b) \in T^{n+1}$ encrypted a message $\mu \in \mathbb{T}$ with noise parameter $\alpha$. 

In the TFHE scheme, the homomorphic evaluation of a binary gate is achieved with operations between TLWE samples and a gate-bootstrapping just after that (except the bootstrapping NOT gate, which does not need bootstrapping). By using this approach, all the basic gates can be evaluated with a single bootstrapping process (GB):

\begin{itemize}
	\item $\mathsf{TFHE.BootsNAND}(c_1, c_2)=\mathsf{GB}((0, \frac{5}{8}) - c_1 - c_2)$
	\item $\mathsf{TFHE.BootsAND}(c_1, c_2)=\mathsf{GB}((0, -\frac{1}{8}) + c_1 + c_2)$
	\item $\mathsf{TFHE.BootsOR}(c_1, c_2)=\mathsf{GB}((0, \frac{1}{8}) + c_1 + c_2)$
	\item $\mathsf{TFHE.BootsXOR}(c_1, c_2)=\mathsf{GB}(2 \cdot (c_1 - c_2))$
	\item $\mathsf{TFHE.NOT}(c)=(0, \frac{1}{4}) - c$
\end{itemize}

The TFHE scheme has the advantages of fast bootstrapping, efficient homomorphic logic circuit evaluation, and so on. Its multi-key version, named MKTFHE~\cite{MKTFHE}, was constructed by Chen et al. in 2019. MKTFHE is the first attempt in the literature to implement an MKFHE scheme in codes.

In the MKTFHE scheme, the ciphertext length increases linearly with the number of users, and a homomorphic NAND gate with bootstrapping is given. The MKTFHE scheme is comprised of the following algorithms.

\begin{itemize}
	\item $\mathsf{MKTFHE.Setup}(1^{\lambda})$: The cloud equipment take a security parameter $\lambda$ as input, and output the public parameter set $mkparams$.
	
	\item $\mathsf{MKTFHE.KeyGen}(mkparams)$: Each party generates its keys independently. First sample the TLWE secret key $sk_i$. Then the algorithm set public key $pk_i$, bootstrapping key $BK_i$ and key-switching key $KS_i$.
	
	\item $\mathsf{MKTFHE.SymEnc}(\mu)$: This algorithm encrypts an input bit $\mu \in \{0, 1\}$, and returns a TLWE ciphertext with the scaling factor $\frac{1}{4}$. The output ciphertext $c = (b, a) \in \mathbb{T}^{n + 1}$ satisfies $b + <a, s> \approx \frac{1}{4} \mu$.
	
	\item $\mathsf{MKTFHE.SymDec}(c, \{ sk_i \})$: 	Input a TLWE ciphertext $c = (b, a_i, ..., a_k)$ and a set of secret keys $\{ sk_i \}$, and return the message $\mu$ which minimizes $|b+\sum_{i = 1}^{k}<a_i, s_i> - \frac{1}{4} m|$.
	
	\item $\mathsf{MKTFHE.MKKeySwitch}(c, \{ KS_i \}_{i \in [k]})$: Input an expa-nded ciphertext $c$ corresponding to $t = (t_1, ... , t_k)$ and a sequence of key-switching keys from $t_i$ to $s_i$ and returns an encryption of the same message under $s = (s_1, ..., s_k)$.
	
	\item $\mathsf{MKTFHE.NAND}(c_1, c_2, \{ pk_i, BK_i, KS_i \}_{i \in [k]})$: Take two TLWE ciphertext as input. This algorithm first extends two input ciphertexts and evaluate the NAND gate homomorphically on encrypted bits. Then the algorithm evaluates the decryption circuit of the TLWE ciphertext and run the $\mathsf{MKTFHE.MKKeySwitch}(c, \{ KS_i \}_{i \in [k]})$ algorithm. Finally, the algorithm outputs a TLWE ciphertext encrypting $m = m_1 \barwedge m_2$.
\end{itemize}

\subsection{Complement operators}

Assume that there are two $k$-bit inputs $a_{in}$ and $b_{in}$, a $k$-bit complement array integer adder can be constructed in following steps.

\begin{enumerate}
	\item Construct a semi-adder with two XOR gates: Input two addends $a$ and $b$ and the carry $c$, return $out = (c \bigoplus (a \bigoplus b))$.
	\item Construct a carrier with three AND gates and two OR gates: Input two addends $a$ and $b$ and the carry $c$, return $c = (a \wedge b ) \vee (a \wedge c ) \vee (b \wedge c)$.
	\item Construct a $1$-bit adder with a semi-adder and a carrier: Input two addends $a$ and $b$ and the carry $c$, return $out = (c \bigoplus (a \bigoplus b))$ and $c = (a \wedge b ) \vee (a \wedge c ) \vee (b \wedge c)$.
	\item Construct a $k$-bit complement array integer adder with k $1$-bit adder: Input two addends $a_{in}$ and $b_{in}$, set the carry $c[0] = 0$, compute $out[i] = (c[i] \bigoplus (a_{in}[i] \bigoplus b_{in}[i]))$, $cout[i + 1] = (a_{in}[i] \wedge b_{in}[i] ) \vee (a_{in}[i] \wedge c[i] ) \vee (b_{in}[i] \wedge c[i])$ with $i = \{ 0, 1, ..., k - 1 \}$. Return the result $out$.
\end{enumerate}

\begin{figure}
	\centering
	\label{fig1}
	\includegraphics[width=1.0\linewidth]{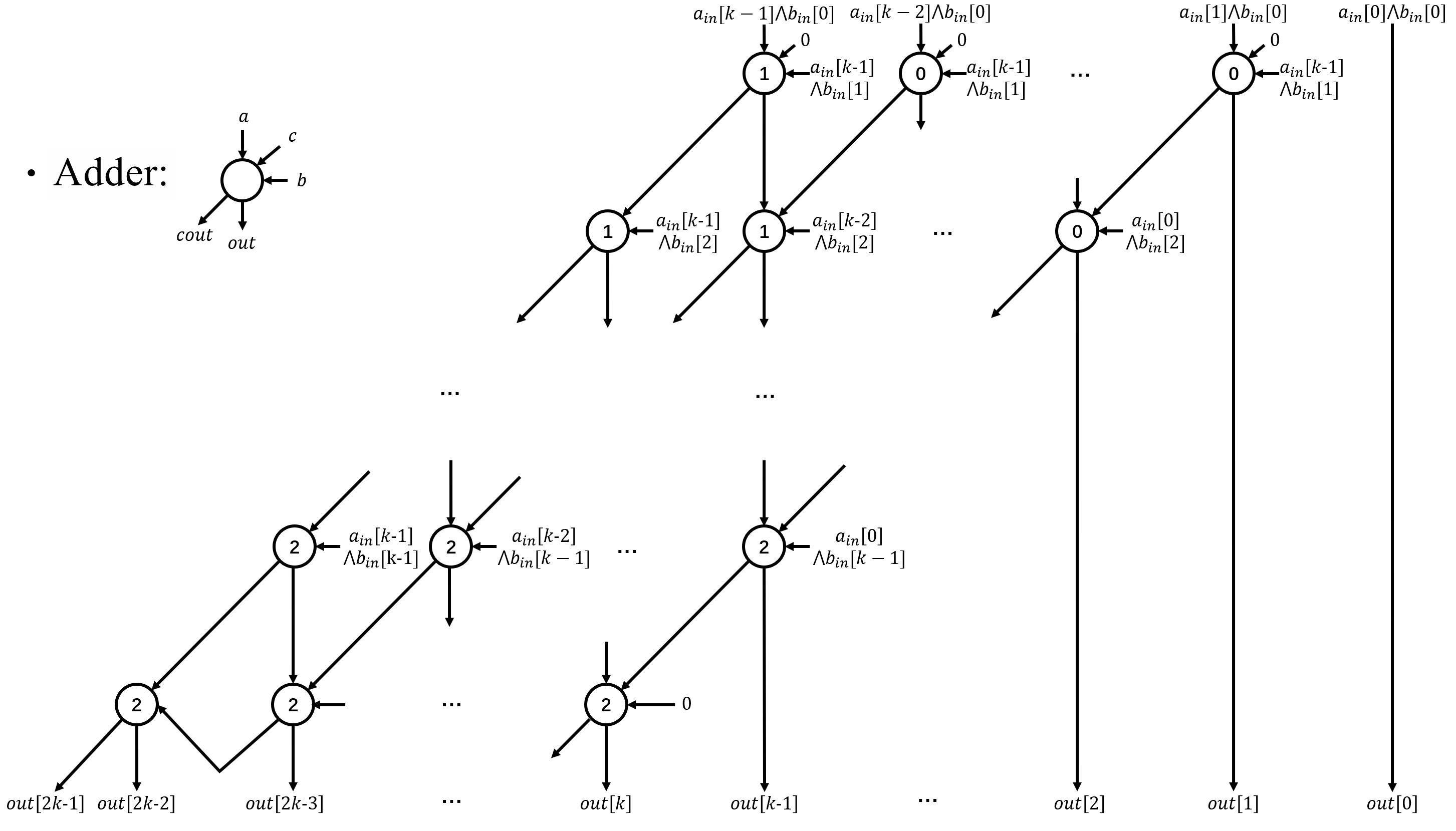}
	\caption{The construction of $k$-bit complement array integer multiplier.The index of the adder stands for the type of it (0-adder, $1$-adder or 2-adder).}
\end{figure}

Assume that there are two $k$-bit inputs $a_{in}$ and $b_{in}$, a $k$-bit complement array integer multiplier can be constructed in following steps.

\begin{enumerate}
	
	\item Prepare three kinds of $1$-bit adder for complement multiplier:
	
	\begin{enumerate}
		
		\item 0-adder: The same as $1$-bit adder above.
		
		\item $1$-adder: Input two addends $a$ and $b$ and the carry $c$, return $out = \overline{(c \bigoplus (\overline{a} \bigoplus b))}$ and $out = (\overline{a} \wedge b ) \vee (\overline{a} \wedge c ) \vee (b \wedge c)$.
		
		\item 2-adder: Input two addends $a$ and $b$ and the carry $c$, return $out = (c \bigoplus (\overline{a} \bigoplus \overline{b}))$ and $out = \overline{(\overline{a} \wedge \overline{b} ) \vee (\overline{a} \wedge c )}$ $\overline{ \vee (\overline{b} \wedge c)}$.
		
	\end{enumerate}
	
	\item Construct a $k$-bit complement array integer multiplier according to the following rules, as shown in Figure 1.
	
	\begin{enumerate}
		
		\item Arrange the adders in $k$ rows and $k - 1$ columns. The input $a$ and $b$ of the adder in $i$ row and $j$ column are $a_{in}[j] \wedge b_{in}[i - 1]$ and $a_{in}[j - 1] \wedge b_{in}[i]$. The inputs $c$ of the adders are $0$ in the first line while the $cout$ of adders above in other lines.
		
		\item The highest bit of each addend (that is to say, $a_{in}[k - 1]$ and $b_{in}[k - 1]$) has the weight $-1$.
		
		\item If two input addends have no weight then use 0-adder.
		
		\item If one of the input addends has the weight $-1$ after AND gate or $1$-adder then use $1$-adder.
		
		\item If one of the addends is an output of a $1$-adder and the other addend has the weight $-1$ after AND gate, use 2-adder.
		
		\item If one of the addends is an output of a 2-adder, use 2-adder.
	\end{enumerate}
\end{enumerate}

Obviously a subtractor can be constructed by an adder and a multiplier. Assume that there are two $k$-bit inputs $a_{in}$ and $b_{in}$, a $k$-bit complement array integer subtractor can be calculated by: $out = a_in + (-1) \times b_in$.

Assume that there are two $k$-bit inputs $a_{in}$ and $b_{in}$, a $k$-bit complement array integer divider~\cite{Vergos2007} can be constructed in following steps.

\begin{enumerate}
	\item Prepare the Controlled Adder/subtractor (CAS) for the divider. We use three bootstrapped XOR gates, two bootstrapped OR gates and two bootstrapped AND gates to construct the CAS. The CAS takes $a_{in}[i]$, $b_{in}[i]$, $c_{in}[i]$ and $p$ as input, and output $out[i]$ and the carrier $c[i + 1]$. The construction of a CAS is shown in Figure 2. The CAS will perform addition if $p = 0$ and perform subtraction if $p = 1$.
	
	\begin{figure}[htb]
		\centering
		\label{fig2}
		\includegraphics[width=1.0\linewidth]{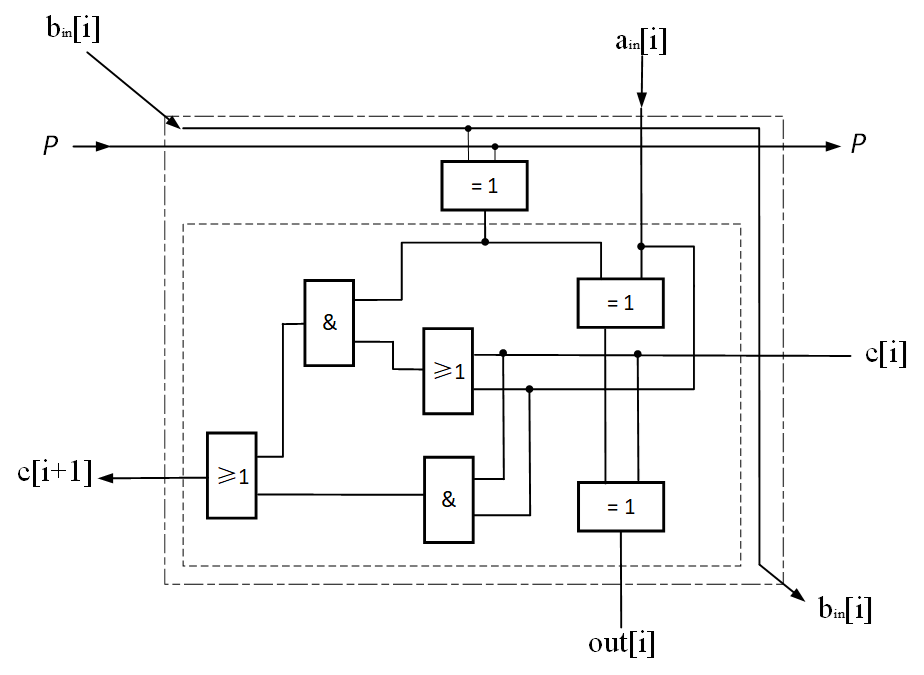}
		\caption{The construction of CAS}
	\end{figure}
	
	\item Then design a absolute value array divider with the CAS above. It takes the $2k$-bit dividend $a_{in}$ and the $k$-bit divisor $b_{in}$ as input, and output the quotient $q$ and the remainder $r$.
	
	\item XOR gates are used to decide the sign bit of the quotient to realize the division of both positive and negative integers.
	
	\item In order to facilitate the operation, the input and output of the divider is unified to complement format. As a result, a compensation device is needed while processing the input and the output. To realize the complement, we first perform XOR gates on the sign bit and each bit, and then add the sign bit to the result, and finally get the complement code.
	
	\item Construct the $k$-bit complement array divider with a absolute value array divider, a XOR gate and two compensation device. The $k$-bit complement array divider takes the $2k$-bit complement dividend $a_{in}$ and the $k$-bit complement divisor $b_{in}$ as input, and output the complement quotient $q$ and the complement remainder $r$. The structure of the $k$-bit complement array divider is shown in Figure 3. For simplicity, we set $k = 3$in this figure.
\end{enumerate}

\begin{figure}[htb]
	\centering
	\label{fig3}
	\includegraphics[width=1.0\linewidth]{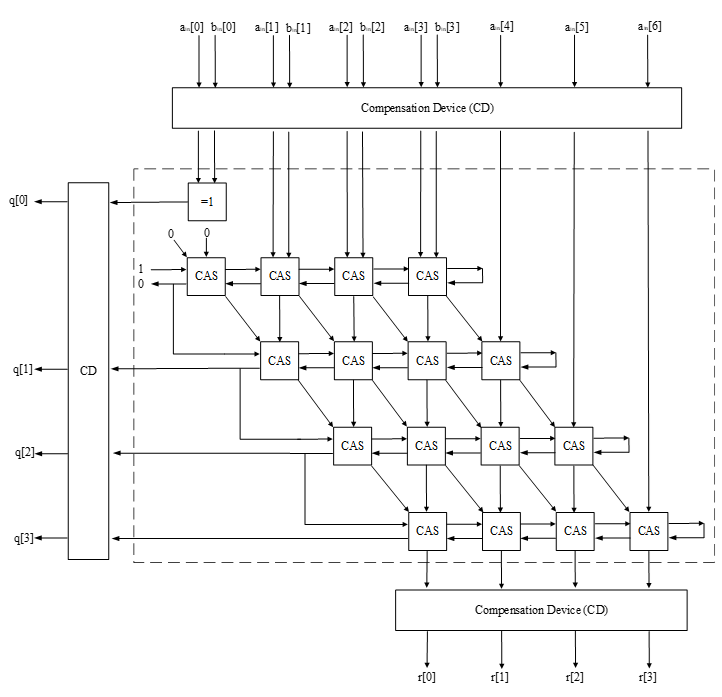}
	\caption{The construction of $k$-bit complement array divider}
\end{figure}

\subsection{Linear regression}

In unary linear regression, there is a calculation formula to obtain the global optimal solution during training the model. According to the definition of linear regression, we can evaluate the accuracy of the model. Suppose we observe $n$ data pairs and call them $\{(x_i, y_i), i = 1, ..., n\}$. We implement unary linear regression and utilize its calculation formula to train model:

\begin{equation}
	(\omega ^* , b ^*) = argmin_{(\omega, b)} \sum_{i = 1}^{m}(f(x_i) - y_i)^2
\end{equation}

By deriving the above formula and making the derivative zero, we can get the calculation formula.

\begin{equation}
	\omega = \frac{\sum_{i = 1}^{m} y_i(x_i - \bar{x})}{\sum_{i = 1}^{m} x_i^2 - \frac{1}{m} (\sum_{i = 1}^{m} x_i)^2 }, \bar{x} = \frac{1}{m}\sum_{i = 1}^{m} x_i
\end{equation}

\begin{equation}
	b = \frac{1}{m} \sum_{i=1}^{m} (y_i - \omega x_i)
\end{equation}

Where $\omega$ is the slope of the line and $b$ is the y-intercept. $\bar{x}$ stands for the average of  $x_1, x_2, \cdots, x_m$.

In most linear regression, there is often no calculation formula to help us obtain the global optimal solution, but the common practice is to utilize the Gradient descent (GD) method to obtain the local optimal through multiple iterations. The iterative formula of the classical gradient descent method is as follows.

\begin{equation}
	\theta_j^{t + 1} = \theta_j^t - \alpha \frac{\partial J(\theta_j^t)}{\theta_j^t}
\end{equation}

\begin{equation}
	J(\theta) = \frac{1}{2m} \sum_{i=1}^{m} (h_{\theta}(x_i) - y_i)^2
\end{equation}

Where $\alpha$ is the learning rate, $t$ is the number of iterations, {$\theta$} is the parameter of model, $J(\theta) $ is the gradient of {$\theta$}, $j$ is the index of parameters to be solved, $m$ is the number of training data and $h_{\theta}$ stands for the prediction model.

\section{The proposed privacy-preserving distributed machine learning}
\label{3Scheme}
Although MKTFHE scheme provides bootstrapped NAND gates, it is time-consuming to use NAND gates directly for privacy computing on Industrial Internet. We expand NAND gates to other fundamental gates to better support mathematical operators. At the same time, we construct an adder, a subtractor, a multiplier and a divider, so that users can perform mathematical operators infinitely.

\subsection{Homomorphic fundamental gates}

In previous works, the other fundamental gates (except the NAND gate) are constructed by joining the multi-key bootstrapped NAND gates together. Although only the multi-key bootstrapped NAND gate is described in ~\cite{Hwang79}, any arbitrary binary bootstrapped gate can be evaluated in the same way. Our scheme uses a similar approach in MKTFHE to directly evaluate the following basic gates in a multi-key version.

\begin{itemize}
	
	\item $\mathsf{MKBootsAND}(c_1, c_2, \{ pk_i \}_{i \in [k]})$: Given ciphertexts $c_1$ and $c_2$, this algorithm first extends two input ciphertexts into $c_1^{'}$ and $c_2^{'}$ under the same key. Then it evaluates the AND gate by computing:
	\begin{equation}
		c^{'} = (0, - \frac{1}{8}) + c_1 + c_2
	\end{equation}
	homomorphically on encrypted bits. Then this algorithm evaluates the decryption circuit of the ciphertext $c^{'}$ to bootstrap it. Finally, this algorithm run the $\mathsf{MKTFHE.MKKey}$ $\mathsf{Switch(c, {KS_i}_{i \in [k]})}$ algorithm and outputs a TLWE ciphertext encrypting $m = m_1 \wedge m_2$.
	
	\item $\mathsf{MKBootsOR}(c_1, c_2, \{ pk_i \}_{i \in [k]})$: Given ciphertexts $c_1$ and $c_2$, this algorithm first extends two input ciphertexts into $c_1^{'}$ and $c_2^{'}$ and evaluates the OR gate by computing:
	\begin{equation}
		c^{'} = (0, \frac{1}{8}) + c_1 + c_2
	\end{equation}
	homomorphically on encrypted bits. Then this algorithm bootstraps it, run the $\mathsf{MKTFHE.MKKeySwitch(c, {KS_i}_{i \in [k]})}$ algorithm and outputs a TLWE ciphertext encrypting $m = m_1 \vee m_2$.
	
	\item $\mathsf{MKNOT}(c, mkparams)$: Take ciphertext $c$ as input, this algorithm evaluates the NOT gate by computing:
	\begin{equation}
		c^{'} = (0, \frac{1}{4}) - c
	\end{equation}
	homomorphically. This computation will not increase the noise in ciphertext or change the key so that no bootstrapping and key-switching process is needed.
	
	\item $\mathsf{MKBootsNOR}(c_1, c_2, \{ pk_i \}_{i \in [k]})$: Given ciphertexts $c_1$ and $c_2$, this algorithm first extends two input ciphertexts into $c_1^{'}$ and $c_2^{'}$ and evaluates the OR gate by computing:
	\begin{equation}
		c^{'} = (0, \frac{1}{8}) - c_1 - c_2
	\end{equation}
	homomorphically on encrypted bits. Then this algorithm bootstraps it, run the $\mathsf{MKTFHE.MKKeySwitch(c, {KS_i}_{i \in [k]})}$ algorithm and outputs a TLWE ciphertext encrypting $m = m_1 \bar{\vee} m_2$.
	
	\item $\mathsf{MKBootsXOR}(c_1, c_2, \{ pk_i \}_{i \in [k]})$: Given ciphertexts $c_1$ and $c_2$, this algorithm first extends two input ciphertexts and evaluates the XOR gate by computing:
	\begin{equation}
		c^{'} = 2 \cdot (c_1 - c_2)
	\end{equation}
	homomorphically. Then run the bootstrapping and key-switching algorithm, and outputs a TLWE ciphertext encrypting $m = m_1 \bigoplus m_2$.
	
	\item $\mathsf{MKBootsXNOR}(c_1, c_2, \{ pk_i \}_{i \in [k]})$: Given ciphertexts $c_1$ and $c_2$, this algorithm first extends two input ciphertexts and evaluates the XOR gate by computing:
	\begin{equation}
		c^{'} = \frac{1}{4} - 2 \cdot (c_1 - c_2)
	\end{equation}
	homomorphically. Then run the bootstrapping and key-switching algorithm, and output a TLWE ciphertext encrypting $m = m_1 \bigodot m_2$.
	
\end{itemize}

First of all, a mapping function $ModToT$ is required to transform a message bit $m$ to an element on torus, like $(0, \frac{1}{8})$ in Equation (6), such that all of the following computations are executed on torus. The goal of Equation (6) is to compute $m_1 \wedge m_2$. Let us look into it more in detail. Then call $LweSub$ to calculate $c^{'}$ by subtraction. Similarly in Equation (7), after mapping message element to torus element, $LweAdd$ is called to calculate $c^{'}$. As shown in Equation (10), $LweSub$ and $LweMul$ should be called. $NOT$ is the simplest operation which only needs subtraction of $LweSub$. Luckily, $ModToT$ and $LweSub$ have been provided in the library of MKTFHE. We have to design and implement two algorithms $LweAdd$ and $LweMul$ by ourselves, to achieve addition and multiplication of two TLWE ciphertexts, independently.

\subsection{Two important components used in bootstrapped gates}

This subsection illustrates two components used in bootstrapped gates, $LweAdd$ and $LweMul$, which are missed in ~\cite{MKTFHE}. The idea of $LweAdd$ is to execute addition on the $n+1$-dimension vectors $c_1$ and $c_2$ element by element. Remember to refresh the current variance by adding the variance of ciphertext $c_1$ and the variance of ciphertext $c_2$ to get the variance of ciphertext $c$, as shown in Algorithm 1. As shown in Equation (4), $LweMul$ is only required in $XOR$. Such multiplication is dot multiplication, which can be expressed by addend $\{ a_1 \}_{i, j}$ plus $k$-time $\{ a_2 \}_{i, j}$. Remember to refresh the current variance by adding the variance of ciphertext $c_1$ and the variance of ciphertext $c_2$ times $k^2$ to get the variance of ciphertext $c$, as shown in Algorithm 2.

\begin{table*}[htb]
	\centering
	\label{Algorithm 1}
	\resizebox{\linewidth}{!}{
	\begin{tabular}{rl}
		\hline
		\multicolumn{2}{l}{\textbf{Algorithm 1} Multi-key TLWE ciphertexts addition ($LweAdd$)}\\
		\hline
		\textbf{Input}:&MKTFHE parameter set, number of the participants $p$ and two MKTLWE ciphertexts\\
		&$c_1 = (a_1, b_1)$ and $c_2 = (a_2, b_2)$\\
		\textbf{Output}:&A MKTLWE ciphertext $c = (a, b) = c_1 + c_2$\\
		1.&Extract the dimension of the lattice $n$ in parameter set.\\
		2.&For $i = 1$ to $p$ do\\
		3.&$\quad$For $j = 1$ to $n$ do\\
		4.&$\quad\quad$Compute $\{ a \}_{i, j} = \{ a_1 \}_{i, j} + \{ a_2 \}_{i, j}$\\
		5.&$\quad$End for\\
		6.&End for\\
		7.&Compute $b = b_1 + b_2$\\
		8.&Refresh the current variance\\
		\hline
	\end{tabular}}
\end{table*}

\begin{table*}[htb]
	\centering
	\label{Algorithm 2}
	\resizebox{\linewidth}{!}{
	\begin{tabular}{rl}
		\hline
		\multicolumn{2}{l}{\textbf{Algorithm 2} Multi-key TLWE ciphertexts multiplication ($LweMul$)}\\
		\hline
		\textbf{Input}:&MKTFHE parameter set, number of the participants $p$, the coefficient $k$ and two MKTLWE \\
		&ciphertexts $c_1 = (a_1, b_1)$ and $c_2 = (a_2, b_2)$\\
		\textbf{Output}:&A MKTLWE ciphertext $c = (a, b) = c_1 + k \cdot c_2$\\
		1.&Extract the dimension of the lattice $n$ in parameter set.\\
		2.&For $i = 1$ to $p$ do\\
		3.&$\quad$For $j = 1$ to $n$ do\\
		4.&$\quad\quad$Compute $\{ a \}_{i, j} = \{ a_1 \}_{i, j} + k \cdot \{ a_2 \}_{i, j}$\\
		5.&$\quad$End for\\
		6.&End for\\
		7.&Compute $b = b_1 + k \cdot b_2$\\
		8.&Refresh the current variance\\
		\hline
	\end{tabular}}
\end{table*}

However, two components used in bootstrapped gates above, $LweAdd$ and $LweMul$, which are missed in ~\cite{MKTFHE}. The idea of $LweAdd$ is to execute addition on the $n+1$-dimension vectors $c_1$ and $c_2$ element by element. Remember to refresh the current variance by adding the variance of ciphertext $c_1$ and the variance of ciphertext $c_2$ to get the variance of ciphertext $c$, as shown in Algorithm 1. As shown in Equation (4), $LweMul$ is only required in $XOR$. Such multiplication is dot multiplication, which can be expressed by addend $\{ a_1 \}_{i, j}$ plus $k$-time $\{ a_2 \}_{i, j}$. Remember to refresh the current variance by adding the variance of ciphertext $c_1$ and the variance of ciphertext $c_2$ times $k^2$ to get the variance of ciphertext $c$, as shown in Algorithm 2.

We summarize the operations between two LWE ciphertexts used in each basic binary gate in Table 1.

\begin{table}[htb]
	\centering
	\label{tab1}
	\caption{Operations used in each basic gate in our scheme}
	\begin{tabular}{|c|c|c|c|c|c|}
		\hline
		Gate&$BS$&$Add$&$Sub$&$Mul$&$ModToT$\\
		\hline
		AND&$\surd$&$\surd$&&&$\surd$\\
		OR&$\surd$&$\surd$&&&$\surd$\\
		NOT&&&$\surd$&&$\surd$\\
		NAND&$\surd$&&$\surd$&&$\surd$\\
		NOR&$\surd$&&$\surd$&&$\surd$\\
		XOR&$\surd$&&$\surd$&$\surd$&\\
		XNOR&$\surd$&&$\surd$&$\surd$&$\surd$\\
		\hline
	\end{tabular}
	
\end{table}

\subsection{
    \texorpdfstring{$k$-bit homomorphic adder}{{} bit homomorphic adder}
    }

After achieving various basic gate operations in subsection 4.1, it is still a gap between such gates and complicated machine learning functions or private computing directly. In this subsection, we construct a $k$-bit complement array integer adder based on the MKTFHE scheme. We choose complement representation instead of true form or inverse code, for the complement can compute both positive and negative numbers in the same way. A $k$-bit complement array integer adder can be constructed in the following four steps. The main idea was described in subsection 3.3. The only difference is that bootstrapping $BS$ is necessary after each computation to decrease noise. The $k$-bit complement array integer adder is shown in Algorithm 3.

\begin{table*}[htb]
	\centering
	\label{Algorithm 3}
	\resizebox{\linewidth}{!}{
	\begin{tabular}{rl}
		\hline
		\multicolumn{2}{l}{\textbf{Algorithm 3} $k$-bit complement array integer adder}\\
		\hline
		\textbf{Input}:&MKTFHE parameter set, two ciphertexts $c_1 = \mathsf{MKTF}$- $\mathsf{HE.SymEnc}(\mu_1)$ and \\
		&$c_2 = \mathsf{MKTFHE.SymEnc}(\mu_2)$ and the public keys of all participants\\
		\textbf{Output}:&Ciphertext $c = \mathsf{MKTFHE.SymEnc}(\mu_1 + \mu_2)$\\
		1.&Set the first carry $c_c[0] = 0$\\
		2.&Encrypt addends and carry with public keys\\
		3.&For $i = 1$ to $k - 1$ do\\
		4.&$\quad$Compute the $out[i]$ and $c_c[i]$\\
		5.&End for\\
		\hline
	\end{tabular}}
\end{table*}

\begin{enumerate}
	\item Construct a homomorphic semi-adder with two bootstrapped XOR gates: Input two ciphertexts $c_a$ and $c_b$ and the encrypted carry $c_c$, return $out = BS(c_c \bigoplus(BS(c_a \bigoplus c_b)))$, where $BS$ stands for bootstrapping process.
	
	\item Construct a homomorphic carrier with three bootstrapped AND gates and two bootstrapped OR gates: Input two ciphertexts $c_a$ and $c_b$ and the encrypted carry $c_c$, return $cout = BS( BS( BS( c_a \wedge c_b) \vee BS(c_a \wedge c_c) ) \vee BS(c_b \wedge c_c) )$.
	
	\item Construct a homomorphic $1$-bit adder by merging a homomorphic semi-adder and a homomorphic carrier: Input two ciphertexts $c_a$ and $c_b$ and the encrypted carry $c_c$, return $out = BS(c_c \bigoplus(BS(c_a \bigoplus c_b)))$ and $cout = BS( BS( BS( c_a \wedge c_b) \vee BS(c_a \wedge c_c) ) \vee BS(c_b \wedge c_c) )$.
	
	\item Construct a homomorphic $k$-bit complement array integer adder with k homomorphic $1$-bit adder: Input two ciphertexts $c_a$ and $c_b$, set the encrypted carry $c_c[0] = 0$, compute $out[i] = BS(c_c[i] \bigoplus(BS(c_a[i] \bigoplus c_b[i])))$ and $c_c[i+1] = BS( BS( BS( c_a[i] \wedge c_b[i]) \vee BS(c_a[i] \wedge c_c[i]) ) \vee BS(c_b[i] \wedge c_c[i]) )$, where $i = \{ 0, 1, ..., $k$-1\}$. Return the result $out$.
\end{enumerate}

In $k$-bit complement array integer adder, the output carry of the last $1$-bit adder is abandoned because of the capture of the complement.

\begin{figure}[htb]
	\centering
	\label{fig4}
	\includegraphics[width=1.0\linewidth]{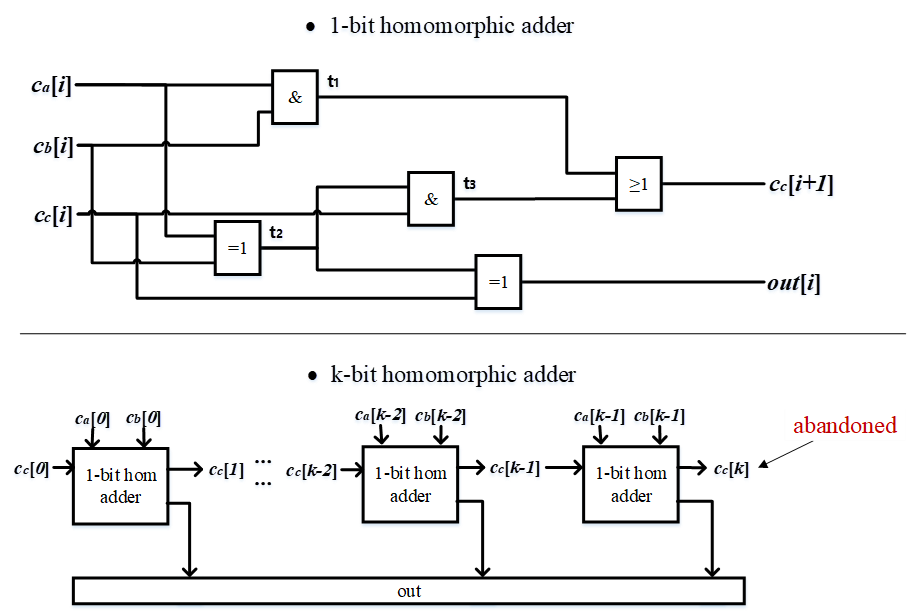}
	\caption{The construction of $k$-bit homomorphic complement array integer adder. GB stands for gate-bootstrapping.}
\end{figure}

\subsection{
    \texorpdfstring{$k$-bit homomorphic multiplier}{{}-bit homomorphic multiplier}
}

We also construct a $k$-bit complement array integer multiplier based on the MKTFHE scheme. A $k$-bit complement array integer multiplier can be constructed in the following steps. The main idea was described in subsection 3.3. The only difference is that bootstrapping $BS$ is necessary after each computation to decrease noise. A $k$-bit complement array integer multiplier is shown in Algorithm 4.

\begin{table*}[htb]
	\centering
	\label{Algorithm 4}
	\resizebox{\linewidth}{!}{
	\begin{tabular}{rl}
		\hline
		\multicolumn{2}{l}{\textbf{Algorithm 4} $k$-bit complement array integer multiplier}\\
		\hline
		\textbf{Input}:&MKTFHE parameter set, two ciphertexts $c_1 = \mathsf{MKTF}$-$\mathsf{HE.SymEnc}(\mu_1)$ and \\
		&$c_2 = \mathsf{MKTFHE.SymEnc}(\mu_2)$ and the public keys of all participants.\\
		\textbf{Output}:&A ciphertext $c = \mathsf{MKTFHE.SymEnc}(\mu_1 \times \mu_2)$\\
		1.&Set the first carry $c_c[0] = 0$\\
		2.&Encrypt numbers and carry with public keys\\
		3.&For $i = 1$ to $k - 2$ do\\
		4.&$\quad$For $j = 0$ to $k - 2$ do\\
		5.&$\quad$$\quad$Compute the AND gate.\\
		6.&$\quad$$\quad$If $j = 0$, compute the 2-homadder\\
		7.&$\quad$$\quad$Elseif $i + j \geq k - 1$, compute the $1$-homadder\\
		8.&$\quad$$\quad$Else compute the 0-homadder\\
		9.&$\quad$End for\\
		10.&$\quad$Get the $(j + k)$th bit of the result\\
		11.&End for\\
		12.&For $j = 0$ to $k - 1$ do \\
		13.&$\quad$Compute the $j$th bit of the output by 2-homadder.\\
		14.&End for\\
		\hline
	\end{tabular}}
\end{table*}

\begin{enumerate}
	
	\item Prepare three kinds of homomorphic $1$-bit adder:
	
	\begin{enumerate}
		
		\item 0-homadder: The same as the homomorphic $1$-bit adder above.
		
		\item $1$-homadder: Input two ciphertexts $c_a$ and $c_b$ and the encrypted carry $c_c$, return $out = \overline{BS(c_c \bigoplus(BS(\overline{c_a} \bigoplus}$ $\overline{c_b)))}$ and $c_{out} = BS( BS( BS( \overline{c_a} \wedge c_b) \vee BS(\overline{c_a} \wedge c_c) ) \vee BS(c_b \wedge c_c) )$.
		
		\item 2-homadder: Input two ciphertexts $c_a$ and $c_b$ and the encrypted carry $c_c$, return $out = BS(c_c \bigoplus(BS(\overline{c_a} \bigoplus$ $\overline{c_b})))$ and $c_{out}=\overline{BS( BS( BS( \overline{c_a} \wedge }$ $\overline{\overline{c_b}) \vee BS(\overline{c_a} \wedge c_c) )}$ $\overline{\vee BS(\overline{c_b} \wedge c_c) )}$.
		
	\end{enumerate}
	
	\item Construct a $k$-bit homomorphic complement array integer multiplier according to the rules in Section 3.4 but use homomorphic adder instead of the previous adder.
	
\end{enumerate}

We use Algorithm 4 to evaluate $k$-bit complement array integer multiplier homomorphically, where $k$ is the number of bits of the addends. In $k$-bit complement array integer multiplier, the output carry of the last $1$-bit adder is considered as the highest bit of the result.

\subsection{
    \texorpdfstring{$k$-bit  homomorphic subtractor and divider}{{}-bit  homomorphic subtractor and divider}
}
We construct a $k$-bit complement array integer subtractor by using a $k$-bit complement array integer adder and a $k$-bit complement array integer multiplier. A $k$-bit complement array integer subtractor is shown in Algorithm 5.

We also construct a $k$-bit complement array integer divider based on the MKTFHE scheme. The main idea was described in subsection 3.3. The only difference is that bootstrapping $BS$ is necessary after each computation to decrease noise. A $k$-bit complement array integer divider can be constructed in the following steps:

\begin{enumerate}
	
	\item Prepare the homomorphic CAS for the divider. We use three homomorphic bootstrapped XOR gates, two homomorphic bootstrapped OR gates and two homomorphic bootstrapped AND gates to construct the homomorphic CAS. The homomorphic CAS takes ciphertexts $c_a[i]$, $c_b[i]$, $c_c[i]$ and $c_p$ as input, and output $c_{out}[i]$ and the carrier $c_{c}[i + 1]$ satisfying $c_{out}[i] = BS(c_a[i] \bigoplus BS( BS( c_b[i] \bigoplus c_p ) \bigoplus c_c[i] ) )$ and $c_{c}[i + 1] = BS( BS( BS(c_a[i] \vee c_c[i]) \wedge BS(c_b[i] \bigoplus c_p) ) \vee BS(c_a[i] \wedge c_c[i]) )$. The homomorphic CAS will perform addition if the message of $c_p$ is $0$ and perform subtraction if the message of $c_p$ is $1$.
	
	\item Then design the homomorphic absolute value array divider with the homomorphic CAS above. It takes the $2k$-bit dividend $c_a$ and the $k$-bit divisor $c_b$ as input, and output the quotient $c_q$ and the remainder $c_r$.
	
	\item Homomorphic XOR gates are used to decide the sign bit of the ciphertext of quotient to realize the homomorphic division of both positive and negative integers.
	
	\item A homomorphic compensation device is also designed to realize the complement division. 
	
	\item Construct the $k$-bit homomorphic complement array divider with a homomorphic absolute value array divider, a homomorphic XOR gate and two homomorphic compensation device. The $k$-bit homomorphic complement array divider takes the ciphertexts of $2k$-bit complement dividend $a_{in}$ and the $k$-bit complement divisor $b_{in}$ as input, and output the ciphertexts of complement quotient $q$ and the complement remainder $r$.
	
\end{enumerate}

\begin{table*}[htb]
	\centering
	\label{Algorithm 5}
	\resizebox{\linewidth}{!}{
	\begin{tabular}{rl}
		\hline
		\multicolumn{2}{l}{\textbf{Algorithm 5} $k$-bit complement array integer subtractor}\\
		\hline
		\textbf{Input}:&MKTFHE parameter set, two ciphertexts $c_1 = \mathsf{MKTF}$-$\mathsf{HE.SymEnc}(\mu_1)$ and \\
		&$c_2 = \mathsf{MKTFHE.SymEnc}(\mu_2)$ and the public keys of all participants.\\
		\textbf{Output}:&A ciphertext $c = \mathsf{MKTFHE.SymEnc}(\mu_1 - \mu_2)$\\
		1.&Change every bit of $c_2$ and add $1$ to calculate $-c_2$\\
		2.&Set the first carry $c_c[0] = 0$\\
		3.&Encrypt addends and carry with public keys\\
		4.&For $i = 1$ to $k - 1$ do\\
		5.&$\quad$Compute the $out[i]$ and $c_c[i]$\\
		6.&End for\\
		\hline
	\end{tabular}}
\end{table*}

\begin{table*}[htb]
	\centering
	\label{Algorithm 6}
	\resizebox{\linewidth}{!}{
	\begin{tabular}{rl}
		\hline
		\multicolumn{2}{l}{\textbf{Algorithm 6} $k$-bit complement array integer divider}\\
		\hline
		\textbf{Input}:&MKTFHE parameter set, two ciphertexts $c_1 = \mathsf{MKTF}$-$\mathsf{HE.SymEnc}(\mu_1)$ and \\
		&$c_2 = \mathsf{MKTFHE.SymEnc}(\mu_2)$ and the public keys of all participants.\\
		\textbf{Output}:&Two ciphertext $q = \mathsf{MKTFHE.SymEnc}(\mu_1 / \mu_2)$ and $q = \mathsf{MKTFHE.SymEnc}(\mu_1 mod \mu_2)$\\
		1.&Encrypt the $p_0$ with public keys\\
		2.&Change the inputs into complement format with compensation device.\\
		3.&Calculate the first later of divider.\\
		4.&For $i = 1$ to $k - 1$ do\\
		5.&$\quad$For $j = 0$ to $k - 1$ do\\
		5.&$\quad$$\quad$Compute the CAS unit.\\
		6.&$\quad$End for\\
		7.&End for\\
		\hline
	\end{tabular}}
\end{table*}

We use Algorithm 6 to evaluate $k$-bit complement array integer divider homomorphically, where $k$ is the number of bits of the divisor. 

In summary, this section first expands NAND gates to other six basic gates, AND, OR, NOT, NOR, XOR and XNOR. In order to compute these gates, two important components, $LweAdd$ and $LweMul$ are then proposed. At last, we construct the adder, subtractor, multiplier and divider. With these homomorphic operators, we can evaluate arbitrary polynomials without decryption.

\subsection{Multi-key homomorphic distributed linear regression scheme}

We first implement a multi-key fully homomorphic encryption scheme in which the secret key is distributed among the parties, while the corresponding collective public key $pk$ is known to all of them. Thus, each party can independently compute on ciphertexts encrypted under $pk$ but all parties have to collaborate to decrypt a ciphertext. This enables the participants to train a collectively encrypted model, that cannot be decrypted as long as one participant is honest and refuses to participate in the decryption. Then we implement two methods: calculation formula method and GD method to train and evaluate the linear regression model.

Calculation formula method: The advantage of this method is that the global optimal solution can be obtained without multiple iterative operation which can reduce the computational cost of time and space. We first estimate the size of the input data, and then select homomorphic operators with appropriate bits to calculate the above formula. Since our operators are integer operators, we choose to directly round off the input data under the condition of ensuring accuracy.
We use algorithm 7 to train the linear regression model under multi-key homomorphic encryption.

GD method: Considering that our homomorphic operators only support integer calculation, and the learning rate $\alpha$ is usually a floating-point number less than 1, we rewrite the GD iterative formula in Subsection 3.4 to zoom the $\alpha$ to integers. We bring the linear regression model into the origin GD iterative formula, the zooming multiple is $n$, and the practical integer GD iterative formula is below:

\begin{equation}
	b' = b \cdot n - \alpha \cdot n \cdot \frac{2}{m} \cdot \sum_{i = 1}^{m} [y_i \cdot n - ( \frac{\omega}{n} \cdot x_i - \frac{b}{n} )]
\end{equation}

\begin{equation}
	\omega' = \omega \cdot n - \alpha \cdot n \cdot \frac{2}{m} \cdot \sum_{i = 1}^{m} [y_i \cdot n - ( \frac{\omega}{n} \cdot x_i - \frac{b}{n} )]
\end{equation}

\begin{equation}
	loss = \frac{1}{m} \cdot \sum_{i = 1}^{m} [y_i - (\omega \cdot x_i + b) / n]^2
\end{equation}

After we get the rewritten iterative formula, we estimate the size of the input data to select the zooming multiple, iterative times and homomorphic operators with appropriate numbers of bits.

We use Algorithm 8 to train the linear regression model under multi-key homomorphic encryption.

\begin{table*}[htb]
	\centering
	\label{Algorithm 7}
	\resizebox{\linewidth}{!}{
	\begin{tabular}{rl}
		\hline
		\multicolumn{2}{l}{\textbf{Algorithm 7} Calculation formula method in multi-key fully homomorphic linear regression}\\
		\hline
		\textbf{Input}:&MKTFHE parameter set, two ciphertexts $c_{x_n} = \mathsf{MKTF}$- $\mathsf{HE.SymEnc}(x_n)$ and \\
		&$c_{y_n} = \mathsf{MKTFHE.SymEnc}(y_n)$ and the public keys of all participants\\
		\textbf{Output}:&Two ciphertext $c_{\omega} = \mathsf{MKTFHE.SymEnc}(\omega)$ and $c_b = \mathsf{MKTFHE.SymEnc}(b)$\\
		1.&Reconstruct the public keys to bootstrapping key\\
		2.&Extend the sing-key ciphertext to multi-key ciphertext\\
		3.&Calculate the average value $c_{\bar{x}}$ of $c_{x}$, following the Equation (1)\\
		4.&Calculate the $c_{\omega}$ by $c_{\bar{x}}$, following the Equation (2)\\
		5.&Calculate the $c_{b}$ by $c_{\omega}$, following the Equation (3)\\
		\hline
	\end{tabular}}
\end{table*}

\begin{table*}[htb]
	\centering
	\label{Algorithm 8}
	\resizebox{\linewidth}{!}{
	\begin{tabular}{rl}
		\hline
		\multicolumn{2}{l}{\textbf{Algorithm 8} GD method in multi-key fully homomorphic linear regression}\\
		\hline
		\textbf{Input}:&MKTFHE parameter set, zooming multiple $n$, iterative times $d$, two sets of ciphertexts\\
		&$c_{x_n} = \mathsf{MKTF}$- $\mathsf{HE.SymEnc}(x_n)$ and $c_{y_n} = \mathsf{MKTFHE.SymEnc}(y_n)$ and the public keys of all\\
		&participants\\
		\textbf{Output}:&Two ciphertext $c_{\omega} = \mathsf{MKTFHE.SymEnc}(\omega)$ and $c_b = \mathsf{MKTFHE.SymEnc}(b)$\\
		1.&Reconstruct the public keys to bootstrapping key\\
		2.&Extend the sing-key ciphertext to multi-key ciphertext\\
		3.&Given iteration initial value to $\omega$ and $b$\\
		4.&For $i = 1$ to $d$ do\\
		5.&$\quad$Calculate the GD iterative equation (13) (14)\\
		6.&$\quad$Set $\omega = \omega '$ and $b = b'$\\
		7.&End for\\
		\hline
	\end{tabular}}
\end{table*}

\section{Implementation and experiments}
\label{4Imp}
The test environment is Ubuntu 18.04 operation system, with Intel Xeon Gold 5220 CPU and 46512MiB memory. 

\subsection{Experiments on homomorphic basic gates}

The MKTFHE scheme provides only homomorphic bootstrapped NAND gates. Our scheme constructs other basic homomorphic bootstrapped gates in the same way as the bootstrapped NAND gate in MKTFHE. We compared our basic bootstrapped gates with gates built by the NAND gates in the same test environment. Experiment shows that our basic bootstrapped gates are as efficient as the bootstrapped NAND gate in the MKTFHE scheme, which is much more efficient than the gates with the same function constructed by only bootstrapped NAND gates.

\begin{table}[htb]
	\centering
	\label{tab2}
	\caption{Expression of our basic bootstrapped gates and the average time of each gate in both our scheme and naïve scheme.}
	\begin{tabular}{|c|c|c|}
		\hline
		Gate&Naïve scheme(s)&Our scheme(s)\\
		\hline
		AND&0.238671&0.238008\\
		OR&0.584701&0.23558\\
		NOT&0.236541&2e-06\\
		NAND&0.235327&0.239638\\
		NOR&0.470561&0.236153\\
		XOR&0.71246&0.2354\\
		XNOR&0.710777&0.23588\\
		\hline
	\end{tabular}
	
\end{table}

The result in Table 2 shows that our scheme is more efficient than the naïve scheme, which connects the bootstrapped NAND gates to realize the operations. We have improved in efficiency, especially in bootstrapped OR gate, bootstrapped NOR gate, NOT gate, bootstrapped XOR gate and bootstrapped XNOR gate: the time cost by a single bootstrapped XOR gate reduced by about 67\% while the NOT gate is evaluated almost instantaneously for no bootstrapping is needed.

\subsection{
    \texorpdfstring{Experiments on $k$-bit homomorphic adder}{Experiments on {}-bit homomorphic adder}
}

We construct a $k$-bit homomorphic complement integer adder based on the basic bootstrapped gates we designed above. We create two participants for our experiment. Each of them creates its addend and secret key individually. For simple, we set $k = 1, 2, ..., 8$.

The server input multi-key parameters, the ciphertexts encrypted under different keys by two participants, and the public keys of two participants. Then the server extends the ciphertext, evaluates the $k$-bit adder homomorphically, and returns the ciphertext to the participants. The participants decrypt the ciphertext and get the result.

According to the structure of $k$-bit homomorphic adder, while computing the addition two $k$-bit addends, $k$ $1$-bit adder is required. As shown in subsection 4.2, it requires $5$ bootstrapped gates to construct a $1$-bit adder. That is to say, $5k$ basic bootstrapped gates are needed to construct a $k$-bit homomorphic adder. Obviously, the cost of the homomorphic evaluation grows linearly with the number of bootstrapped gates. So, the cost of the adder grows linearly with the bits of input numbers.

\begin{table}[htb]
	\centering
	\label{tab3}
	\caption{The average time of $k$-bit homomorphic adder.}
	\begin{tabular}{|c|c|c|c|}
		\hline
		Bits&Time(s)&Num of gates&Time/gate(s)\\
		\hline
		$1$-bit&1.19515&5&0.23903\\
		$2$-bit&2.39738&10&0.239738\\
		$3$-bit&3.60485&15&0.240323\\
		$4$-bit&4.80614&20&0.240307\\
		$5$-bit&6.01884&25&0.240754\\
		$6$-bit&7.25483&30&0.241828\\
		$7$-bit&8.43608&35&0.241031\\
		$8$-bit&9.69468&40&0.242367\\
		\hline
	\end{tabular}
	
\end{table}

\begin{figure}[htb]
	\centering
	\label{fig5} 
	\includegraphics[width=1.0\linewidth]{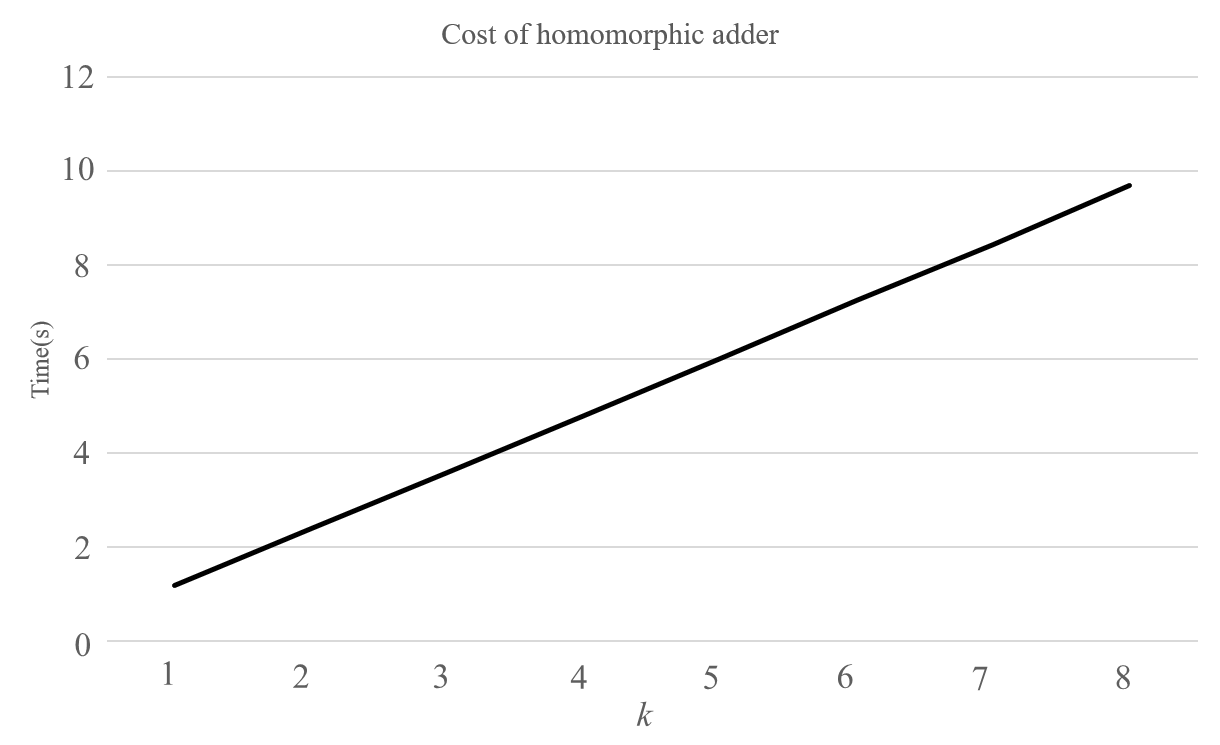}
	\caption{The cost of $k$-bit homomorphic complement array integer adder}
\end{figure}

From Table 3 and Figure 5, we can learn that the number of binary bootstrapped gates grows linearly with the number of the bits of addends. As evaluation is the most expensive step in our scheme, we can say that the cost of the $k$-bit homomorphic complement adder grows linearly with the bits of input numbers.

\subsection{
    \texorpdfstring{Experiments on $k$-bit homomorphic multiplier}{Experiments on {}-bit homomorphic multiplier}
}

To construct a $k$-bit homomorphic multiplier, we construct 3 kinds of homomorphic adder: 0-homadder, 1-homadder and 2-homadder. We also construct a $k$-bit homomorphic complement integer multiplier based on three kinds of adders and the basic bootstrapped gates we designed above. The structure of 3 kinds of homomorphic adder in section 4.3 shows that it requires $7$ basic bootstrapped gates to construct a homomorphic adder: $5$ gates in $1$-bit adder and two AND gates to compute the input. It can be learned from the structure of the $k$-bit homomorphic multiplier in section 3.3 that it requires $k \times (k - 1)$ homomorphic adders to construct a $k$-bit homomorphic multiplier. That is to say, it requires $7 \times k \times (k - 1)$ basic bootstrapped gates for a $k$-bit homomorphic multiplier. So, the cost of the multiplier grows quadratically with the bits of input numbers. Note that there is parallel optimization in the multiplier, so the average time consumed by a gate is much less than that consumed by a single gate.

We create two participants for our experiment, each of them creates its complement number and secret key individually. For simplicity, we set $k = 2, 3, ..., 8$. The server input multi-key parameters, the ciphertexts encrypted under different keys by two participants, and the public keys of two participants. Then the server extends the ciphertext, evaluates the $k$-bit multiplier homomorphically, and returns the ciphertext to the participants. The participants decrypt the ciphertext and get the result.

\begin{table}[htb]
	\centering
	\label{tab4}
	\caption{The average time of each step of experiments on $k$-bit homomorphic adder, as well as the number of gates used in multiplier.}
	\begin{tabular}{|c|c|c|c|}
		\hline
		bit of mul&Time(s)&Gates&Time/gate(s)\\
		\hline
		2-bit mul&1.14368&14&0.081691429\\
		3-bit mul&2.66844&42&0.063534286\\	
		4-bit mul&4.83101&84&0.057512024\\
		5-bit mul&7.43843&140&0.053131643\\
		6-bit mul&10.762&210&0.051247619\\
		7-bit mul&14.6522&294&0.049837415\\
		8-bit mul&19.1052&392&0.048737755\\
		\hline
	\end{tabular}
	
\end{table}

\begin{figure}[htb]
	\centering
	\label{fig6}
	\includegraphics[width=1.0\linewidth]{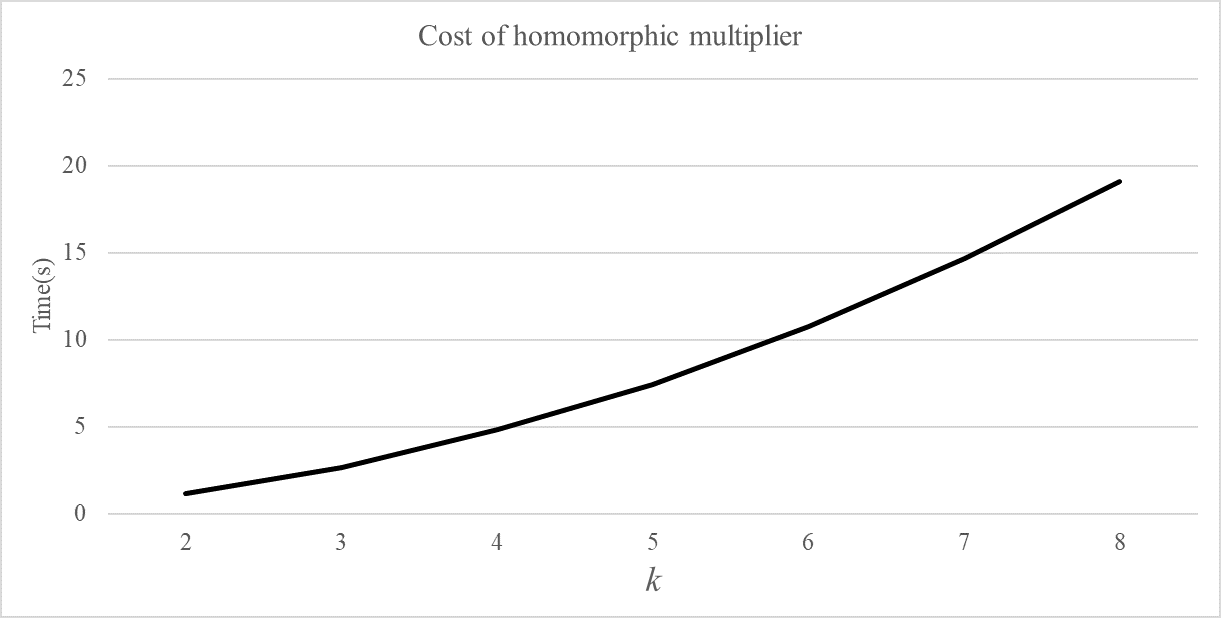}
	\caption{The cost of $k$-bit homomorphic complement array integer multiplier}
\end{figure}

From Table 4 and Figure 6, we can learn that the number of binary bootstrapped gates grows quadratically with the number of the bits of addends. Similarly, we can say that the cost of the $k$-bit homomorphic complement multiplier grows quadratically with the bits of input numbers.

\subsection{
    \texorpdfstring{Experiments on $k$-bithomomorphic subtractor and divider}{Experiments on {}-bit homomorphic subtractor and divider}
}
We construct a $k$-bit homomorphic complement integer subtractor based on the basic bootstrapped gates we designed above. We create two participants for our experiment. Each of them creates its addend and secret key individually. For simple, we set $k = 1, 2, ..., 8$.

According to the structure of $k$-bit homomorphic subtractor, while computing the $k$-bit subtraction, $k$ $1$-bit adder is required. As shown in subsection 4.2, it requires $5$ bootstrapped gates to construct a $1$-bit adder. At the same time, a bootstrapped XOR gate is used to compute the negation. That is to say, $6k$ basic bootstrapped gates are needed to construct a $k$-bit homomorphic adder. Obviously, the cost of the homomorphic evaluation grows linearly with the number of bootstrapped gates. So, the cost of the subtractor grows linearly with the bits of input numbers.

\begin{table}[htb]
	\centering
	\label{tab5}
	\caption{The average time of experiments on $k$-bit homomorphic subtractor}
	\begin{tabular}{|c|c|c|c|}
		\hline
		bit of sub&Time(s)&Gates&Time/gate(s)\\
		\hline
		1&1.99209&6&0.332015\\
		2&2.90969&12&0.242474167\\	
		3&4.45103&18&0.247279444\\
		4&5.84264&24&0.243443333\\
		5&7.42889&30&0.247629667\\
		6&8.93639&36&0.248233056\\
		7&10.35&42&0.246428571\\
		8&11.9013&48&0.24794375\\
		\hline
	\end{tabular}
	
\end{table}

\begin{figure}[htb]
	\centering
	\label{fig7}
	\includegraphics[width=1.0\linewidth]{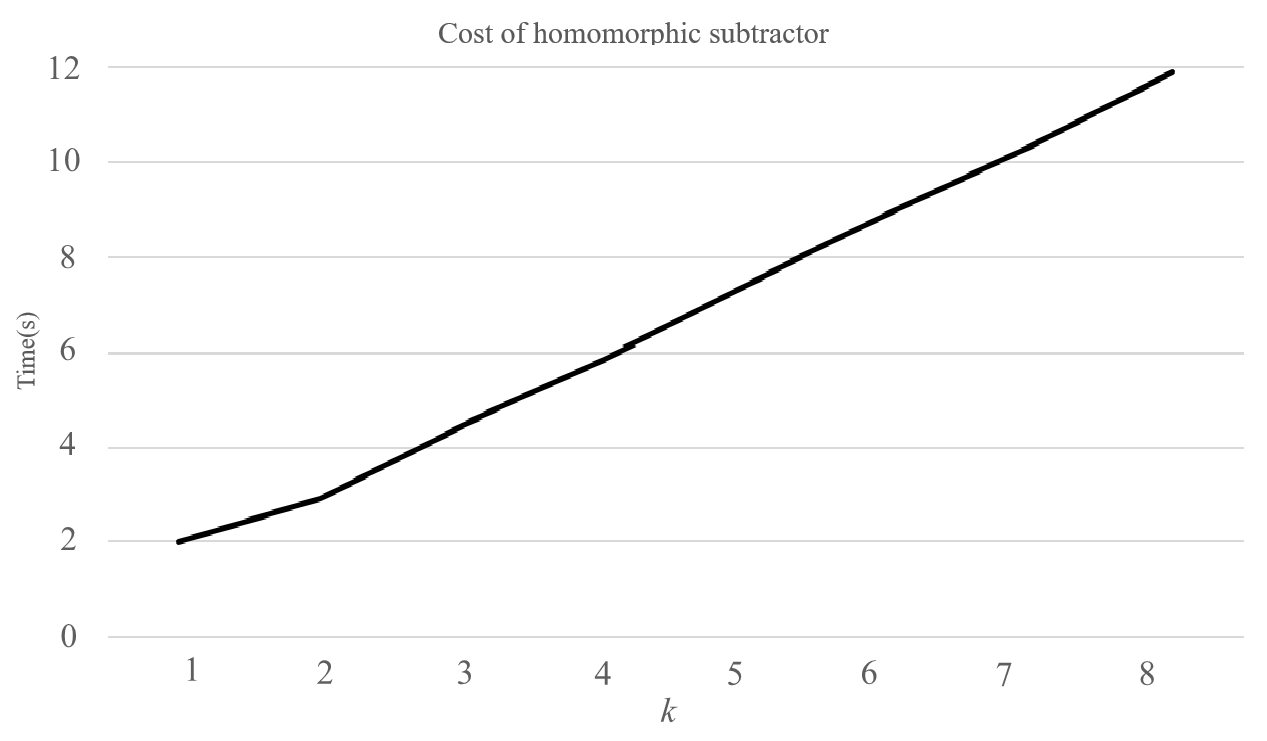}
	\caption{The cost of $k$-bit homomorphic complement array integer subtractor}
\end{figure}

Experiments show that the cost of $k$-bit homomorphic subtractor grows linearly with the bit of minuend.

We also construct a $k$-bit homomorphic complement integer divider in the similar way. We create two participants for our experiment. Each of them creates its addend and secret key individually. For simple, we set $k = 1, 2, ..., 8$.

According to the structure of $k$-bit homomorphic divider, while computing the $k$-bit division, a homomorphic absolute value array divider, a homomorphic XOR gate and two homomorphic compensation device are required. There are $7$ homomorphic gates in each hom-CAS while there are $k^2$ hom-CASs in each homomorphic absolute value array divider. It takes $2k$ homomorphic gates to construct a compensation device. As a result, $7k^2 + 2k + 1$ basic bootstrapped gates are needed to construct a $k$-bit homomorphic divider. Note that there are more optimization in the divider than in the multiplier, and the CAS almost run in parallel, so a $k$-bit divider is faster than a $k$-bit multiplier, and the cost of the divider grows almost linearly with the number of layers.

\begin{table}[htb]
	\centering
	\label{tab6}
	\caption{The average time of experiments on $k$-bit homomorphic divider}
	\begin{tabular}{|c|c|c|c|}
		\hline
		bit of div&Time(s)&Layers&Time/layer(s)\\
		\hline
		1&1.483&1&1.483\\
		2&4.09858&2&2.04929\\
		3&7.08719&3&2.362396667\\
		4&10.0022&4&2.50055\\
		5&13.1937&5&2.63874\\
		6&15.965&6&2.660833333\\
		7&19.1003&7&2.728614286\\
		8&22.2938&8&2.786725\\
		\hline
	\end{tabular}
	
\end{table}

\begin{figure}[htb]
	\centering
	\label{fig8}
	\includegraphics[width=1.0\linewidth]{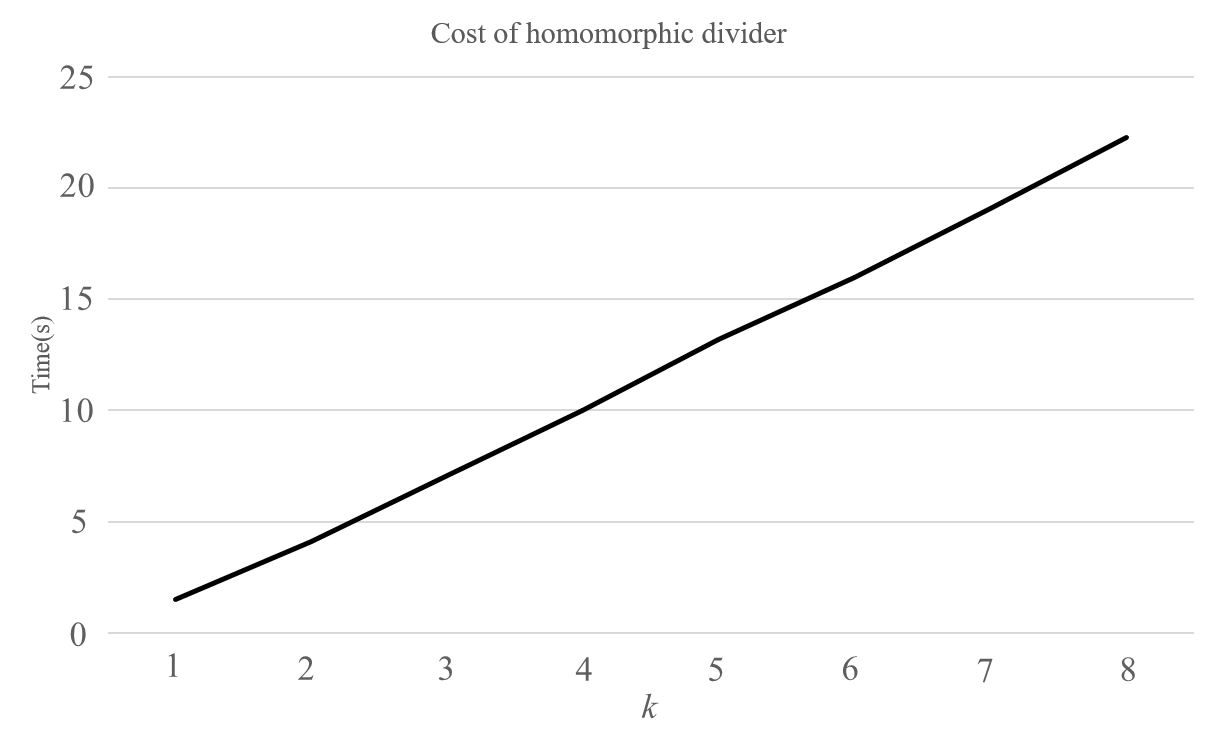}
	\caption{The cost of $k$-bit homomorphic complement array integer divider}
\end{figure}

\subsection{Experiments on multi-key homomorphic distributed linear regression}

We implement a multi-key fully homomorphic encryption scheme in which the secret key is distributed among the parties, while the corresponding collective public key $pk$ is known to all of them. Thus, each party can independently compute on ciphertexts encrypted under $pk$ but all parties have to collaborate to decrypt a ciphertext. This enables the participants to train a collectively encrypted model, that cannot be decrypted as long as one participant is honest and refuses to participate in the decryption. 

Our multi-key fully homomorphic linear regression scheme is composed by multiple parties which can be divided by three types of entities: participants, cloud server, and decryption party, we take two participants as examples. the whole scheme are shown in the Figure 9, the steps are as follows:

\begin{enumerate}
	\item Participants have the need to outsource computing, and each participant offers its own part of data for model training. During the step of data encryption, all participants call $\mathsf{MKTFHE.KeyGen}(mkparams)$ to generate their symmetric key, asymmetric key bootstrapping key and key-switching key independently, then call $\mathsf{MKTFHE.SymEnc}(\mu)$ to encrypt the data. And finally upload the ciphertext of input data and public key to the cloud server.
	
	\item Cloud server is usually composed of one or more high-performance servers, and doesn’t have their own data. After receiving the ciphertext data from participants, the cloud server will use the participants’ key to generate a new bootstrapping key and extend the single-key ciphertext of the input data to multi-key ciphertext. Then cloud server uses our designed homomorphic operators to train the linear regression model. After the training is finished, the result of the model will be sent to the decryption party.
	
	\item When the decryption party receive the ciphertext of the trained model, they will unite all the participants to call $\mathsf{MKTFHE.SymDec}(c, \{sk_i\})$ to decrypt the result.
\end{enumerate}

\begin{figure}[htb]
	\centering
	\label{fig9}
	\includegraphics[width=1.0\linewidth]{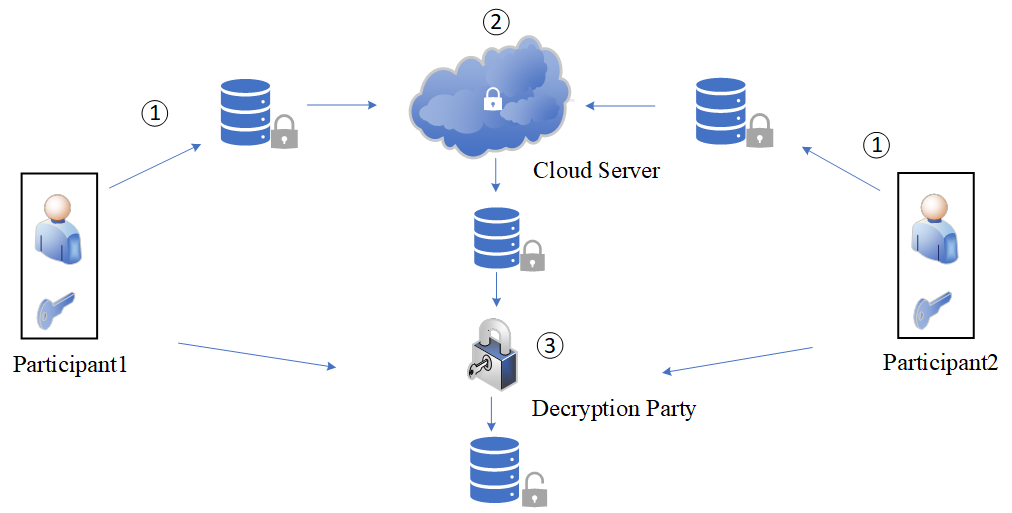}
	\caption{Multi-key fully homomorphic linear regression scheme}
\end{figure}

We show our setting of the parameters in the Table 7, following the notation of MKTFHE library. The achievement estimated security level of our scheme is 110-bit while the dimension of the TLWE problem is $k = 1$.

\begin{table*}[htb]
	\centering
	\label{tab7}
	\caption{Parameter sets of multi-key homomorphic linear regression}
	\resizebox{\linewidth}{!}{
	\begin{tabular}{|c|c|c|c|c|c|c|c|}
		\hline
		LWE-$n$&LWE-$\alpha$&LWE-$B'$&LWE-$d'$&RLWE-$N$&RLWE-$\beta$&RLWE-$B$&RLWE-$d$\\
		\hline
		560&$3.05 \times 10^{-5}$&$2^2$&8&1024&$3.72 \times 10^{-9}$&$2^9$&3\\
		\hline
	\end{tabular}}
	
\end{table*}

We train a multi-key fully homomorphic linear regression model in two methods by using our proposed homomorphic operators. The input data is composed of several sets of linear data and some random noise. Considering the size of the data, we chose 8-bit homomorphic operators to train the model in formula method and 16-bit homomorphic operators in GD method. In addition, the learning rate and the zooming multiple of the GD method is 0.001 and 10,000. We make several experiments and find that the model will converge by performing up to 10 iterations. We also create $m$ participants in our experiment. Each of them generates their own secret key to encrypt their own data. For simple, we set $m = 2,4,8$, and the $3,4,5,6$ columns record the time of each step.

\begin{table*}[htb]
	\centering
	\label{tab8}
	\caption{The result of experiments on multi-key fully homomorphic linear regression}
	\resizebox{\linewidth}{!}{
	\begin{tabular}{|c|c|c|c|c|c|}
		\hline
		method&Num of Parties&KeyGen(s)&Ciphertext extension(s)&Training(s)&Evaluation(s)\\
		\hline
		Formular&$k=2$&$1.984$&$0.0008$&$227.467$&$37.131$\\
		Formular&$k=4$&$4.002$&$0.0016$&$445.593$&$68.968$\\
		Formular&$k=8$&$8.916$&$0.0035$&$869.148$&$138.259$\\
		GD&$k=2$&$2.025$&$0.0008$&$1133.33$/iter&$190.644$\\
		GD&$k=4$&$4.008$&$0.0015$&$2092.17$/iter&$372.353$\\
		\hline
	\end{tabular}}
	
\end{table*}

In Table 8, the result shows that: the running time including key reconstruction time, extension ciphertext time, training time and evaluation time all grow linearly with the numbers of participants; due to the large zooming multiple of the GD method, the homomorphic operators with larger bits are selected. At the same time, the GD method needs multiple iterations, and the calculation process is more complex, so the running time including training time and evaluation time is much more than that of the formula method.

Note that there is more optimization in our linear regression scheme such as running in parallel. Besides, with our homomorphic operators, we can implement more complex multi-key fully homomorphic machine learning scheme.

\section{Conclusion and Discussion}
\label{5Con}
\subsection{Conclusion}

The scheme proposes a series of $k$-bit homomorphic complement operators based on MKTFHE, which narrows the gap between the original NAND gate and complicated machine learning functions. Experiment shows that the cost of the adder and the subtractor grows linearly with the bits of input numbers and the cost of the multiplier grows quadratically. Meanwhile, the cost of the divider grows almost linearly with the number of layers in it.

To narrow the gap between multi-key bootstrapped NAND gates and multi-key homomorphic mathematical operators, we construct other basic bootstrapped gates (AND, OR, NOT, NOR, XOR, and XNOR) in the same way as the bootstrapped NAND gate in MKTFHE and with the same efficiency as the NAND gate. Experiment shows that constructing basic binary gates in this way is much more efficient than building them by directly joining the bootstrapped NAND gates together, especially when constructing XOR gate and NOT gate. Then we construct a $k$-bit complement adder, a $k$-bit complement subtractor, a $k$-bit complement multiplier and a $k$-bit complement divider based on our basic binary bootstrapped gates. Finally, we train distributed linear regression model by utilizing our proposed multi-key homomorphic operators. The operators we designed can be directly used to achieve privacy-preserving distributed machine learning schemes in distributed communication system.


\subsection{Discussion}

The plaintext space of the MKTFHE scheme is $\{0,1\}$, which is smaller than the plaintext space in other homomorphic encryption schemes like BGV~\cite{BrakerskiGV14} or CKKS~\cite{CheonKKS17, CheonHKKS18}. Extending the message space on torus may be helpful. It has been experimented that integers or fixed-point numbers can also be mapped to a ring on the torus~\cite{BouraGGJ20, BouraGG18} as well as performing homomorphic evaluations. If the message space of MKTFHE is magnified to integers or fixed-point numbers, it may reduce the overhead in time and space in MKTFHE.

Common reference strings (CRS) are needed in this scheme for all the participants, and the computing server has to know the multi-key parameters in advance. If the CRS can be removed in MKTFHE, the scheme will no longer need a trusted third party to generate the CRS, making the scheme more secure.

We described only multi-key homomorphic linear regression model in our scheme, but more machine learning schemes (such as logistic regression, SVM, etc) can be evaluated in the same way, for the basic mathematical operators are provided in our scheme.

\section{Acknowledgement}

This work is supported by National Natural Science Foundation of China (No. 61872109), Shenzhen Basic Research Project, China (No. JCYJ20180507183624136), Shenzhen Basic Research Project, China (No. JCYJ20200109113405927) and National Science and Technology Major Project Carried on by Shenzhen, China (No. CJGJZD20200617103000001).

\bibliography{mybibfile}

\end{document}